\documentclass{article}

\usepackage{PRIMEarxiv}
\usepackage{amsmath, amssymb}
\usepackage{algorithmic}
\usepackage{algorithm}
\usepackage{graphicx}
\usepackage{url}
\usepackage{booktabs}

\usepackage{listings}
\usepackage{xcolor}

\definecolor{codegreen}{rgb}{0,0.6,0}
\definecolor{codegray}{rgb}{0.5,0.5,0.5}
\definecolor{codepurple}{rgb}{0.58,0,0.82}
\definecolor{backcolour}{rgb}{0.95,0.95,0.92}

\lstdefinestyle{mystyle}{
	backgroundcolor=\color{backcolour},   
	commentstyle=\color{codegreen},
	keywordstyle=\color{magenta},
	numberstyle=\tiny\color{codegray},
	stringstyle=\color{codepurple},
	basicstyle=\ttfamily\footnotesize,
	breakatwhitespace=false,         
	breaklines=true,                 
	captionpos=b,                    
	keepspaces=true,                 
	numbers=left,                    
	numbersep=5pt,                  
	showspaces=false,                
	showstringspaces=false,
	showtabs=false,                  
	tabsize=2
}
\title{Communication-Efficient and Memory-Aware Parallel Bootstrapping using MPI\thanks{This work was inspired by the author's experience teaching a course on "High-Performance Computing".}}
\author{
	Di Zhang \\
	School of Advanced Technology \\
	Xi'an Jiaotong-Liverpool University \\
	Suzhou, Jiangsu, China \\
	\texttt{di.zhang@xjtlu.edu.cn}
}
\date{}

\begin{document}
	
	\maketitle
	
	\begin{abstract}
		Bootstrapping is a powerful statistical resampling technique for estimating the sampling distribution of an estimator. However, its computational cost becomes prohibitive for large datasets or a high number of resamples. This paper presents a theoretical analysis and design of parallel bootstrapping algorithms using the Message Passing Interface (MPI). We address two key challenges: high communication overhead and memory constraints in distributed environments. We propose two novel strategies: 1) \emph{Local Statistic Aggregation}, which drastically reduces communication by transmitting sufficient statistics instead of full resampled datasets, and 2) \emph{Synchronized Pseudo-Random Number Generation}, which enables distributed resampling when the entire dataset cannot be stored on a single process. We develop analytical models for communication and computation complexity, comparing our methods against naive baseline approaches. Our analysis demonstrates that the proposed methods offer significant reductions in communication volume and memory usage, facilitating scalable parallel bootstrapping on large-scale systems.
	\end{abstract}
	
	\section{Introduction}
	
	Bootstrapping~\cite{efron1979bootstrap} is a cornerstone of modern computational statistics. By repeatedly resampling from the original data with replacement, it allows for estimating the properties of an estimator, such as its variance and confidence intervals, without stringent parametric assumptions. Despite its versatility, the computational demand of bootstrapping is substantial, often requiring thousands of resamples, which makes parallelization essential for practical applications with large datasets.
	
	While parallel computing frameworks like MPI offer a path to acceleration, a straightforward parallelization of the bootstrapping algorithm often leads to significant performance bottlenecks. The primary challenges are:
	\begin{itemize}
		\item \textbf{Communication Overhead}: Naively distributing the data and collecting all resampled datasets can saturate the network, becoming the dominant cost.
		\item \textbf{Memory Constraints}: In a distributed setting, a single process may not have the memory capacity to hold the entire original dataset or all local resamples, rendering simple approaches infeasible.
	\end{itemize}
	
	This paper provides a theoretical treatment of parallel bootstrapping algorithms designed to mitigate these issues. We present and analyze two optimized strategies:
	\begin{enumerate}
		\item A \emph{Local Statistic Aggregation} method that leverages the mathematical definition of variance to minimize data transfer.
		\item A \emph{Synchronized Pseudo-Random Number Generation} scheme for memory-constrained environments where data must be distributed across processes.
	\end{enumerate}
	We develop analytical models for the communication and computation time of these methods and compare them against progressively more sophisticated baseline algorithms. Our theoretical results show that the proposed strategies achieve superior scalability by effectively bounding communication costs and adhering to strict memory limits.
	
	\section{Related Work}
	
	Parallel bootstrapping has been explored in various contexts. Early work often focused on shared-memory architectures or embarrassingly parallel implementations where little communication is required~\cite{efron1994introduction}. For distributed memory systems, a common approach is to broadcast the data and perform resampling independently, collecting the results~\cite{kleiner2014scalable}. However, the communication cost of this collection phase is often not addressed optimally.
	
	More advanced strategies involve tree-based reductions for aggregating statistics~\cite{li2014efficient} or using specialized communication patterns. Our work distinguishes itself by providing a clear theoretical comparison of point-to-point communication strategies under explicit memory constraints, a common scenario in high-performance computing (HPC) environments.
	
	\section{Problem Formulation and Baseline Algorithm}
	
	\subsection{Problem Setup}
	We consider the problem of estimating the variance of the sample mean, $Var(\widetilde{M})$, via bootstrapping. Let $D$ be the size of the original dataset, and $N$ be the number of bootstrap resamples. The goal is to parallelize this computation across $P$ processes using MPI. We assume $N$ is divisible by $P$ for load balancing. The original dataset is an array of $D$ floating-point numbers (4 bytes each). The network bandwidth is $B$ bytes/second, and latency is neglected. Each process has a computational capacity of $S$ sample-points per second.
	
	\subsection{Naive Baseline Algorithm (Data Broadcast \& Sample Return - DBSR)}
	A straightforward parallel algorithm, given in the initial problem statement, works as follows:
	\begin{enumerate}
		\item The root process (Rank 0) broadcasts the entire dataset ($4D$ bytes) to all other $(P-1)$ processes.
		\item Each process, including the root, independently generates $N/P$ bootstrap samples (each of size $D$) from the data.
		\item All non-root processes send their $N/P$ full samples ($4D \cdot N/P$ bytes) back to the root process.
		\item The root process calculates the mean of each sample and then computes the variance of these $N$ means.
	\end{enumerate}
	
	\section{Analytical Modeling}
	
	We now develop analytical models to compare the performance of different parallel bootstrapping strategies. We focus on \textbf{communication time} ($T_{comm} = \text{Data Volume} / B$) and \textbf{memory usage} per process. Computation time is modeled as $T_{comp} = (\text{Total Sample Points}) / S$.
	
	\subsection{Baseline and Proposed Strategies}
	
	\subsubsection{Strategy A: Full Sample Distribution (FSD)}
	This naive strategy generates all samples on the root and distributes them for processing.
	\begin{itemize}
		\item \textbf{Communication:} Root sends $N$ samples: $4DN$ bytes. Results are negligible. $T_{comm} \approx 4DN / B$.
		\item \textbf{Memory (Root):} Must store all $N$ samples: $O(DN)$.
		\item \textbf{Analysis:} This is impractical for large $N$ or $D$ due to prohibitive memory and communication costs.
	\end{itemize}
	
	\subsubsection{Strategy B: Data Broadcast \& Sample Return (DBSR) - Naive Baseline}
	This is the algorithm described in Section 3.2.
	\begin{itemize}
		\item \textbf{Communication:}
		\begin{align*}
			\text{Data Broadcast:} &\quad 4D(P-1) \\
			\text{Sample Return:} &\quad 4D \cdot (N/P) \cdot (P-1) \\
			T_{comm} &= \frac{4D(P-1)(1 + N/P)}{B}
		\end{align*}
		\item \textbf{Memory:} Each process stores the original data ($D$) and its local samples ($DN/P$), i.e., $O(D + DN/P)$.
		\item \textbf{Analysis:} Communication scales with $O(DN)$, which becomes a bottleneck for large $N$.
	\end{itemize}
	
	\subsubsection{Strategy C: Data Broadcast \& Statistic Aggregation (DBSA) - Proposed Optimization 1}
	We improve Strategy B by having processes compute and return summary statistics.
	\begin{itemize}
		\item \textbf{Algorithm:} Each process computes the mean for each of its $N/P$ bootstrap samples. It then calculates the mean of these means ($m_1$) and the mean of their squares ($m_2$). Only these two values are sent back to the root. The root computes $Var(\widetilde{M}) = m_2 - m_1^2$.
		\item \textbf{Communication:}
		\begin{align*}
			\text{Data Broadcast:} &\quad 4D(P-1) \\
			\text{Statistic Return:} &\quad 4 \times 2 \times (P-1) = 8(P-1) \\
			T_{comm} &= \frac{4D(P-1) + 8(P-1)}{B}
		\end{align*}
		\item \textbf{Memory:} Same as B, $O(D + DN/P)$.
		\item \textbf{Analysis:} This is a major improvement. The $O(DN)$ term is eliminated from communication, which now scales only with $O(D)$. This makes the method highly efficient for a large number of resamples $N$.
	\end{itemize}
	
	\subsubsection{Strategy D: Distributed Data \& RNG Synchronization (DDRS) - Proposed Optimization 2}
	This strategy is designed for memory-constrained environments where no process can hold the entire dataset.
	\begin{itemize}
		\item \textbf{Assumption:} The original data is distributed. Process $i$ holds a distinct segment of the data of size $D/P$. Memory per process is capped at $O(D/P)$.
		\item \textbf{Algorithm:} All processes use an identical pseudo-random number seed. For each bootstrap sample, a global index $j$ ($0 \le j < D$) is generated. Process $i$ is responsible for contributing the data point from its local segment if $j$ falls within its range. Each process computes a partial sum for the sample. The root collects these partial sums from all processes for each of the $N$ samples to calculate the global sample mean.
		\item \textbf{Communication:} For each of the $N$ bootstrap samples, $(P-1)$ processes send one partial sum (1 float) to the root.
		\begin{align*}
			T_{comm} &= \frac{4 \times 1 \times (P-1) \times N}{B} = \frac{4N(P-1)}{B}
		\end{align*}
		\item \textbf{Memory:} Each process stores only $D/P$ original data points, satisfying the memory constraint.
		\item \textbf{Computation:} $T_{comp} = (N \times D) / S$. Although $P$ checks are performed per sample point, the sampling operation itself is distributed.
		\item \textbf{Analysis:} This method trades increased communication ($O(NP)$) compared to Strategy C for minimal, bounded memory usage. It is the only feasible option when the $O(D)$ memory requirement of other strategies cannot be met.
	\end{itemize}
	
	\subsection{Theoretical Comparison Summary}
	
	Table~\ref{tab:comparison} summarizes the theoretical performance of the different strategies, focusing on the dominant terms for large $N$ and $D$.
	
	\begin{table*}[t]
		\centering
		\caption{Theoretical comparison of parallel bootstrapping strategies.}
		\label{tab:comparison}
		\begin{tabular}{l l l l}
			\toprule
			\textbf{Strategy} & \textbf{Communication Complexity} & \textbf{Memory per Process} & \textbf{Primary Use Case} \\
			\midrule
			A: FSD        & $O(DN)$ & $O(DN)$ (root) & Impractical \\
			B: DBSR       & $O(DN)$ & $O(D + DN/P)$ & Small $D$, small $N$ \\
			\textbf{C: DBSA} & $\mathbf{O(D)}$ & $O(D + DN/P)$ & \textbf{General purpose, large $N$} \\
			\textbf{D: DDRS} & $\mathbf{O(NP)}$ & $\mathbf{O(D/P)}$ & \textbf{Memory-constrained, large $D$} \\
			\bottomrule
		\end{tabular}
	\end{table*}
	
	\textbf{Key Insights:}
	\begin{itemize}
		\item \textbf{Strategy C (DBSA)} provides a dramatic improvement over the naive baseline (B) by reducing communication complexity from $O(DN)$ to $O(D)$. This makes it the preferred choice for most scenarios where memory is not a limiting factor.
		\item \textbf{Strategy D (DDRS)} is essential for memory-constrained environments. Its $O(NP)$ communication cost is higher than C's $O(D)$ but is independent of $D$, making it suitable for very large datasets. Its memory usage is optimal, scaling as $1/P$.
	\end{itemize}
	
	\section{Algorithm Details}
	
	\subsection{Local Statistic Aggregation (Strategy C)}
	
	The core improvement is the \texttt{summary} function and the modified communication logic on the root and worker processes. The mathematical basis is the identity $Var(X) = E[X^2] - E[X]^2$.
	
	\begin{lstlisting}[language=Python, caption=MPI Code for Local Statistic Aggregation (Strategy C)]
		from mpi4py import MPI
		import numpy as np
		
		comm = MPI.COMM_WORLD
		R = comm.Get_rank()
		P = comm.Get_size()
		
		N = 1000  # Number of bootstraps
		D = 10000 # Dataset size
		
		def bootstrap(data):
		# ... generates N/P bootstrap samples from data ...
		pass
		
		def summary(samples):
		means = np.mean(samples, axis=1)
		return np.array([np.mean(means), np.mean(means**2)])
		
		if R == 0:
		data = np.random.randn(D)  # Generate data on root
		# Send data to all workers
		for i in range(1, P):
		comm.send(data, dest=i)
		# Compute local statistics on root
		local_samples = bootstrap(data)
		stats = summary(local_samples)
		# Receive statistics from workers
		for i in range(1, P):
		worker_stats = comm.recv(source=i)
		stats = np.vstack((stats, worker_stats))
		# Calculate final variance
		global_m1 = np.mean(stats[:, 0])
		global_m2 = np.mean(stats[:, 1])
		variance_of_mean = global_m2 - global_m1**2
		print(f"Variance of the sample mean: {variance_of_mean}")
		else:
		data = comm.recv(source=0)
		local_samples = bootstrap(data)
		stats = summary(local_samples)
		comm.send(stats, dest=0)
	\end{lstlisting}
	
	\subsection{Distributed Resampling with Synchronized RNG (Strategy D)}
	
	This algorithm relies on deterministic pseudo-random number generation to ensure all processes construct the same sequence of global bootstrap indices without communicating them.
	
	\begin{lstlisting}[language=Python, caption=MPI Code for Distributed Resampling (Strategy D)]
		from mpi4py import MPI
		import numpy as np
		
		comm = MPI.COMM_WORLD
		I = comm.Get_rank()
		P = comm.Get_size()
		
		N = 1000
		D = 100000
		assert D % P == 0
		local_D = D // P
		
		# Each process holds its portion of the data
		local_data = np.random.randn(local_D) # In practice, data would be loaded distributedly
		
		np.random.seed(205) # Global seed
		
		def distributed_bootstrap_sample(local_data, I, P, global_D):
		local_sum = 0.0
		local_count = 0
		# Generate global_D indices to form one bootstrap sample
		for _ in range(global_D):
		# Generate a global random index
		global_idx = np.random.randint(0, global_D)
		# Check if this index belongs to my local segment
		if I * local_D <= global_idx < (I + 1) * local_D:
		local_index = global_idx - I * local_D
		local_sum += local_data[local_index]
		local_count += 1
		# Return partial sum and count for this sample
		return np.array([local_sum, local_count])
		
		if I == 0:
		all_means = []
		for n in range(N): # For each bootstrap round
		# Perform local part of sampling
		local_result = distributed_bootstrap_sample(local_data, I, P, D)
		global_sum = local_result[0]
		global_count = local_result[1] # Count should ideally be D, used for verification
		# Collect partial sums from all other processes
		for j in range(1, P):
		remote_result = comm.recv(source=j)
		global_sum += remote_result[0]
		global_count += remote_result[1]
		sample_mean = global_sum / D
		all_means.append(sample_mean)
		variance_of_mean = np.var(all_means)
		print(f"Variance of the sample mean: {variance_of_mean}")
		else:
		for n in range(N):
		local_result = distributed_bootstrap_sample(local_data, I, P, D)
		comm.send(local_result, dest=0)
	\end{lstlisting}
	
	\section{Conclusion and Future Work}
	
	This paper presented a theoretical analysis of communication-efficient and memory-aware parallel bootstrapping algorithms. We proposed two main strategies: Local Statistic Aggregation (DBSA) and Distributed Resampling with Synchronized RNG (DDRS). Through analytical modeling, we demonstrated that DBSA effectively eliminates the dependency on the number of resamples $N$ in communication complexity, while DDRS provides a viable solution for memory-constrained environments at the cost of $O(NP)$ communication.
	
	For future work, it would be valuable to incorporate non-blocking communication to overlap computation and communication further. Exploring dynamic load balancing to handle potential imbalances in the distributed resampling of Strategy D and extending these strategies to heterogeneous computing environments are promising research directions.
	
	\bibliographystyle{unsrt}
	\bibliography{references}
	
\end{document}